\documentclass[sigconf]{acmart}
\usepackage{graphicx} 
\usepackage{booktabs}
\usepackage{graphicx}
\usepackage{subcaption}
\usepackage{mwe}
\usepackage{algorithm}
\usepackage{algpseudocode}
\usepackage{mathtools}
\usepackage{amsmath}
\DeclareMathOperator*{\argmax}{arg\,max}

\def\BibTeX{{\rm B\kern-.05em{\sc i\kern-.025em b}\kern-.08emT\kern-.1667em\lower.7ex\hbox{E}\kern-.125emX}}
    
\copyrightyear{2019}
\acmYear{2019}
\setcopyright{acmlicensed}
\acmConference[ICCAD '19]{ICCAD '19: International Conference on Computer Aided Design}{November 04--07, 2019}{The Westin Westminster
Westminster, Co}
\acmPrice{15.00}
\acmDOI{10.1145/1122445.1122456}
\acmISBN{978-1-4503-9999-9/18/06}

\begin{document}

%
\title[BagNet]{BagNet: Berkeley Analog Generator with Layout Optimizer Boosted with Deep Neural Networks}

%

\author{Kourosh Hakhamaneshi, Nick Werblun, Pieter Abbeel, Vladimir Stojanovi\'{c}}
\email{kourosh_hakhamaneshi, nwerblun, vlada@berkeley.edu}
\email{pabbeel@cs.berkeley.edu}
\affiliation{%
  \institution{University of California Berkeley, USA}
}

%
\renewcommand{\shortauthors}{Hakhamaneshi, et al.}
%
\begin{abstract}
The discrepancy between post-layout and schematic simulation results continues to widen in analog design due in part to the domination of layout parasitics. This paradigm shift is forcing designers to adopt design methodologies that seamlessly integrate layout effects into the standard design flow. Hence, any simulation-based optimization framework should take into account time-consuming post-layout simulation results. This work presents a learning framework that learns to reduce the number of simulations of evolutionary-based combinatorial optimizers, using a DNN that discriminates against generated samples, before running simulations. Using this approach, the discriminator achieves at least two orders of magnitude improvement on sample efficiency for several large circuit examples including an optical link receiver layout.
\end{abstract}

\maketitle

\section{Introduction}
In the integrated circuit (IC) industry, design of analog circuits is still one of the main factors determining development cost and production time. This is mainly due to the fact that analog ICs are usually customized for specific applications and generally lack modularity as opposed to digital design. Thus, most analog ICs are comprised of manually designed blocks composed together to form a complex system. Designing the entire system involves many iterations with human experts exploring the complex multi-dimensional design space and running lengthy simulations to converge on a working solution.
\par
With the emergence of generator-based tools like BAG \cite{chang_bag2:_2018} we can now specify experts' layout design methodology in a parametric way, independent of the underlying technology, that can instantiate manufacturable layouts in different nodes with no extra work. Given this tool, design effort and iteration time can be reduced by creating optimization tools to find the proper parameters for a given performance criteria, based on post-layout simulation results. Here, the optimization tool leverages the initial design-space reduction performed by human experts who embedded their expertise into parameterized layout generators. Simulation of post-layout extracted analog circuits is a major bottleneck of iteration time, and therefore, the optimization tool must be sample efficient. 
\par 
Analog circuit design automation has substantial history in the CAD area, especially with the introduced emphasis on population based optimization methods in the mid 90s \cite{phelps_anaconda:_2000, barros_analog_2010, krasnicki_maelstrom:_nodate}, and even more so recently \cite{prajapati_two_2015}. Unfortunately, population-based methods are, in general, sample inefficient and were primarily used to size small circuits based solely on fast schematic simulations. With recent advancement in technology nodes, layout parasitics have become an increasingly important factor in the design performance degradation, adding disparity between the schematic and the post-layout simulation results that must be considered at design time. Post-layout simulation can be very slow even for simple circuits, mainly because of the required parasitic elements that significantly expand the size of the netlist.
\par
\par
In this regard, Bayesian Optimization (BO) \cite{lyu_efficient_2018} addresses the sample efficiency by 
incorporating uncertainty estimates in an analytical acquisition function. Optimizing this function determines the next sample that maximizes the expected improvement of the main objective. The drawback with this approach is that finding the global optimum of the function itself can still be very expensive. Also, Gaussian processes have limited expressive power, and therefore, might not be a viable modeling choice for complicated design problems. 
\par
In this paper, we combine the evolutionary optimization approaches with Deep Learning methods to boost optimizer's sample efficiency. Specifically, our model learns some information from past experience that can be used to improve the evolution process in the following steps.
\par
In the first section we will detail the implementation of our framework, and in the following sections, we will demonstrate practical experiments to show how it can scale up to handle complex mixed-signal circuits. 

\section{Framework Implementation}
\subsection{Definition of the Problem}
The objective in analog circuit design is usually to minimize one igure of merit (FOM) subject to some hard constraints (strict inequalities). For instance, in op-amp design the objective can be minimizing power subject to gain and bandwidth constraints. However, in practice there is also a budget for metrics in the FOM (i.e power less than 1mW). Therefore, the optimization can be rephrased as a constraint satisfaction problem (CSP) where the variables are circuit's geometric parameters, and outputs are specifications of the circuit topology. Designers can always tighten the budgets to see if there is any other answer with a better FOM that meets their needs.\par
However, in cases where there is no feasible solution to a CSP, designers still prefer to know which solutions are nearly satisfactory to gain insight into which constraint can be adjusted to satisfy their needs.\par
With this in mind we define the following non-negative cost function where finding the zeros is equivalent to finding answers to the CSP problem. If no answer exists, the minimum of this cost and the non-zero terms can give insight about which metrics are the limiting factors:
\begin{equation}
    cost(x) = \sum_{i}{w_ip_i(x)}
    \label{eq: cost}
\end{equation}
where $x$ presents the geometric parameters in the circuit topology and $p_i(x)=\frac{|c_i - c_i^*|}{c_i+c_i^*}$ (normalized spec error) for designs that do not satisfy constraint $c_i^*$, or zero if they do. $c_i$ denotes the value of constraint $i$ at input $x$, and is evaluated using a simulation framework. $c_i^*$ denotes the optimal value. Intuitively this cost function is only accounting for the normalized error from the unsatisfied constraints, and $w_i$ is the tuning factor, determined by the designer, which controls prioritizing one metric over another if the design is infeasible. \par

\subsection{Population Based Methods: Benefits and Drawbacks}
Population based methods have been extensively studied in the past \cite{phelps_anaconda:_2000, barros_analog_2010, krasnicki_maelstrom:_nodate, prajapati_two_2015}, in the application of analog circuit design automation. These methods usually start from an initial population and iteratively derive a new population from the old one using some evolutionary operations (i.e. combination, mutation). Some selection mechanism then picks the elites of the old and the new population for the next generation, a process known as elitism. This process continues until the average cost of the current population reaches a minimum.\par 
While this could work in principle, it is very sample inefficient, and prone to instability in convergence. Therefore, the process must be repeated numerous times due to its stochastic nature. As a result, these methods are not suitable for layout based optimizations where simulation takes a significant amount of time. \par
The sample inefficiency arises from two factors. Firstly, the majority of the new population will only slightly improve upon their ancestors, and as the population improves, the difficulty of replacing old designs increases. Therefore, it would take many iterations until the children evolve enough to surpass the average of the parents. Much of the previous work seeks to improve this by focusing on modifying the evolutionary operations such that they would increase the probability of producing better children, while preserving the diversity \cite{prajapati_two_2015}. Unfortunately, these methods have not been able to sufficiently improve the sample efficiency to accommodate the post-layout simulations. Secondly, many of these methods only look at the total cost value and do not consider sensitivity of the cost to each design constraint, meaning that they do not account for how each specification metric is affecting the overall cost. Expert analog designers usually do this naturally by prioritizing their design objectives depending on what constraint mostly limits their design. Considering only the total cost value can be misleading and may obfuscate useful information about the priority of optimizing the metrics.\par
To address the first issue, if we had access to an oracle which could hypothetically tell us how two designs were compared in terms of each design constraint, we could use it to direct the selection of new designs. Each time a new design is generated we can run the oracle to see how the new design compares to some average design from the previous generation. In circuit design this oracle is in fact the simulator, which is time consuming to query. DNNs seem to be extremely good in approximating complex functions, and generalizing to unseen samples. In this paper we devise a DNN model that can imitate the behaviour of such an oracle. \par
To address the second problem, we can look at the current population and come up with a set of critical specifications (i.e specifications that are the most limiting and should be prioritized first). Each step that we query the oracle, we only add designs that have better performance than the reference design in all metrics in the critical specification set. The important point to note is that once a metric enters the critical specification set it never becomes uncritical, as we do not want to forget which specifications derived the selection of population so far. For finding the critical specification at each time step we use a heuristic which is best described by the pseudo code in algorithm \ref{alg: heuristic}. The intuition behind this heuristic is that the population is sorted by performance in previous critical metrics. Then the specification that results in maximum penalty among all the top designs (i.e. top 10 designs) is chosen as the new critical metric.\par 
\begin{algorithm}
    \begin{algorithmic}
    \caption{Pseudo-code of the heuristic used for updating critical specification list}
    \label{alg: heuristic}
    \State Given population $\mathcal{B}$, specification list $\mathcal{S}$, critical specification list $\mathcal{CS}$ (empty at first), a reference index $k$ (i.e 10)
    \If {$\mathcal{CS}$.empty()}
        \State $\widetilde{\mathcal{B}} \leftarrow$ sort $\mathcal{B}$ by $cost(x)=\sum\limits_{i \in \mathcal{S}} w_i*p_i(x) $
    \Else
        \State $\widetilde{\mathcal{B}} \leftarrow$ sort $\mathcal{B}$ by $cost(x)=\sum\limits_{i \in \mathcal{CS}} w_i*p_i(x) $
    \EndIf
    \State critical\_spec $\leftarrow \argmax\limits_{i \in \mathcal{S}}{\max\limits_{x\in \widetilde{\mathcal{B}}[0:k-1]} {w_i * p_i(x)}}$ 
    \State $\mathcal{CS}.append(\text{critical\_spec})$
    \end{algorithmic}
\end{algorithm}

Using a simulator with the aforementioned heuristics we can decide whether to add a new design to the population or not. With this oracle we can significantly reduce the number of iterations for convergence if we know what designs to add. 
\par
However, we cannot use this oracle if we want to scale our method to do layout-level optimizations on more complex circuit topologies with larger design spaces and more expensive simulation runs. This is because the oracle has to run simulations for all generated instances to determine which designs to add or reject and therefore, there is no real benefit in the number of simulations that it runs. In the next section we propose a DNN structure that can imitate the behavior of the oracle while reducing the required number of simulation runs.

\begin{figure}
    \centering
    \includegraphics[width=\linewidth]{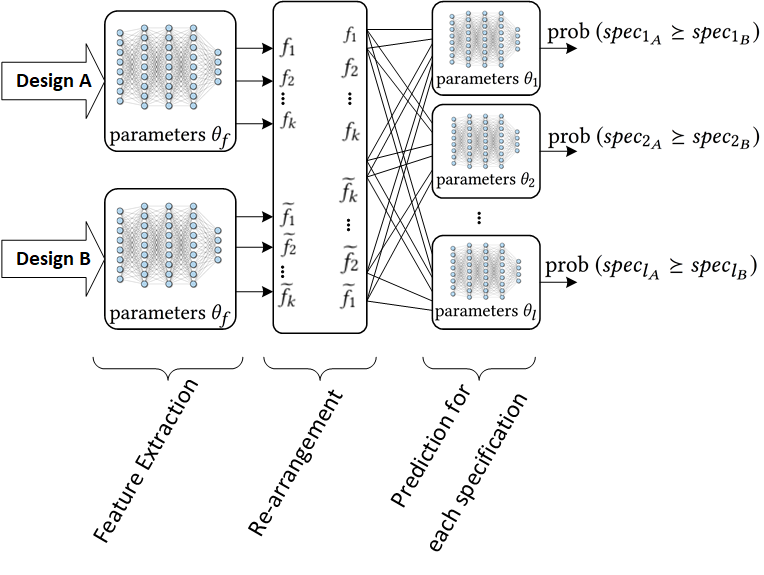}
    \caption{DNN's model used in the system, $\theta=[\theta_f, \theta_1, ..., \theta_l]$ contains the parameters of the DNN. $\mathcal{M}_\theta(D_A, D_B)$ is the output probabilities of the DNN parametrized by $\theta$ for inputs $D_A$ and $D_B$. $\mathcal{M}_\theta(D_A, D_B;i)$ denotes the predicted probability for the $i^{th}$ specification. Note that $\succeq$ denotes preference, not greater than. }
    \label{fig:nn_model}
\end{figure}
\subsection{Model for Imitating the Oracle}
For imitating the oracle there are multiple options. First, we can have a regression model to predict the cost value and then use this predicted value to determine whether or not to accept a design. The cost function that the network tries to approximate can be non-convex and ill-conditioned. Thus, from a limited number of samples it is very unlikely that it would generalize well to unseen data. Moreover, the cost function captures too much information from a single scalar number, so it would be hard to train.\par 
Another option is to predict the value of each metric (i.e. gain, bandwidth, etc.). While the individual metric behaviour can be smoother than the cost function, predicting the actual metric value is unnecessary, since we are simply attempting to predict whether a new design is superior to some other design. Therefore, instead of predicting metric values exactly, the model can take two designs and predict only which design performs better in each individual metric. Figure \ref{fig:nn_model} illustrates the model architecture used for imitating the oracle. 
\par
The inputs to the model are the circuit parameters of two designs that we wish to compare. The model consists of a feature extraction component comprised of only fully connected layers with similar weights for both Design A and Design B. For each specification there is a sub-DNN that predicts the preference over specifications using fully connected layers. There is a subtle constraint that the network should predict complementary probabilities for inputs $[D_A, D_B]$ vs. $[D_B, D_A]$ (i.e. $\mathcal{M}_\theta(D_A, D_B) = 1-\mathcal{M}_\theta(D_B, D_A)$) meaning that there should be no contradiction in the predicted probabilities depending on the order by which the inputs were fed in. To ensure that this property holds and to make the training easier, we can impose this constraint on the weight and bias matrices in the decision networks. To do so, each sub-DNN's layer should have even number of hidden units, and the corresponding weight and bias matrices should be symmetric according to the following equations. \par
\begin{equation*}
    y_{m\times 1} = \textbf{W}_{m\times 2k}x_{2k\times 1}+b_{m\times 1}
\end{equation*}

$$
    \begin{bmatrix}
    y_1\\
    y_2\\
    \vdots\\
    y_m\\
    \end{bmatrix} = 
    \begin{bmatrix}
    \textbf{W}_{\frac{m}{2} \times 2k}\\
    \widetilde{\textbf{W}}_{\frac{m}{2} \times 2k}
    \end{bmatrix}
    \begin{bmatrix}
    x_{k\times 1}\\
    \widetilde{x}_{k \times 1}
    \end{bmatrix}
    +
    \begin{bmatrix}
    b_{\frac{m}{2}\times 1}\\
    \widetilde{b}_{\frac{m}{2}\times 1}
    \end{bmatrix}
$$
Where we have, 
$$
    \begin{aligned}[b]
        & \widetilde{\textbf{W}}(i,j) = \textbf{W}(\frac{m}{2}-1-i, 2k-1-j) \\
        & \text{for } i = 0, ..., \frac{m}{2}-1 \text{ and } j = 0, ..., 2k-1  \\
        & \widetilde{b}(i) = b(\frac{m}{2}-1-i) \text{ for } i = 0, ..., \frac{m}{2}-1
    \end{aligned}
$$
If the weight and bias parameters are set as above, when the input order is changed from $[D_A, D_B]$ to $[D_B, D_A]$ the very first $x$ vector is changed from $[f_1, ..., f_k, \widetilde{f}_k, ..., \widetilde{f}_1]$ to $[\widetilde{f}_1, ..., \widetilde{f}_k, f_k, ..., f_1]$. Thus, for the last layer that has two outputs, the sigmoid function will produce $1-\mathcal{M}_{\theta}(D_A, D_B)$ instead of $\mathcal{M}_{\theta}(D_A, D_B)$.
\par
To train the network, we construct all $(D_A, D_B)$ permutations from the buffer of previously simulated designs and label their comparison in each metric. We then update network parameters with stochastic gradient descent, using sum of cross-entropy loss for all metrics.\par
To avoid over-fitting and being certain about false positive and negatives, we can leverage Bayesian DNNs within our model \cite{neal_bayesian_1996}, which can estimate the uncertainty regarding the decisions. In this paper we use drop out layers which can be considered as Bayesian DNNs with Bernouli distributions \cite{gal_dropout_2015}. During inference we sample the model 5 times and average the probabilities to reduce the uncertainty about the decisions.
For the sake of experiments in this paper, for the feature extractor, we used 2 fully connected (FC) layers of size 20 with RELU non-linearity. For each individual sub DNN after, we used 1 FC layer of size 20 with RELU non-linearity. For output layer we used sigmoid non-linearity to translate scalar logits to probabilities. 
\par
\subsection{Algorithm}
\begin{figure}
    \centering
    \includegraphics[width=\linewidth]{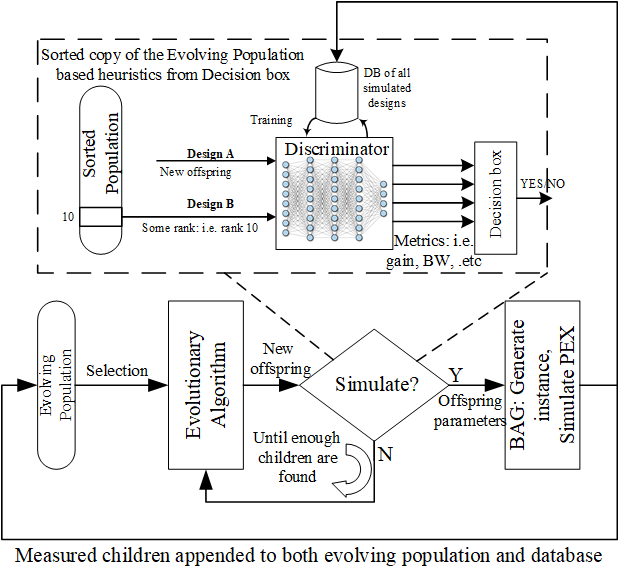}
    \caption{High level architecture of our optimizer}
    \label{fig:flowchart}
\end{figure}

To put everything together, Figure \ref{fig:flowchart} illustrates the high level architecture of the optimizer system. We use the current population and perform some evolutionary operations to get the next generation of the population. The choice of evolutionary operations is somewhat arbitrary as long as they converge, given large enough time and number of samples. We used both cross-entropy [\cite{de_boer_tutorial_2005}] and some canonical $\mu + \lambda $ evolutionary strategies (involving cross-over, blending, and mutation operations) and in this paper we only report the later for the sake of brevity. After generation of new samples we do not simply simulate them and consider them as the next generation parents. We use the DNN to predict if they will be better compared to some reference design already within our current population, and if the answer is positive we simulate the designs, call them the next generation, and proceed with the evolutionary algorithm. \par
The next generation of designs may not have come from the same probability distribution as their parents which the DNN was trained on. To mitigate the distribution drift, for each new approved offspring we run the actual simulator and re-train the DNN model with new data and correct labels. This idea is very similar to DAgger \cite{ross_reduction_nodate} except that we do not relabel all of the children, rather we only relabel the accepted ones. \par
Algorithm \ref{alg: final} shows the entire algorithm, step by step. The algorithm starts off by collecting some random simulated designs and pre-training the DNN. Then, at each iteration, it updates the critical specification list according to the heuristic mentioned earlier and picks a reference design from the population using the critical metrics. Then, for each newly generated design from the evolutionary algorithm, the DNN predicts whether it is superior to the reference design in all critical metrics. The evolutionary algorithm keeps generating designs until enough superior samples are found. After running simulation for those approved samples, we update the population and DNN parameters using SGD. This process is repeated until either we find a solution or a maximum number of iterations is reached. 

\begin{algorithm}
    \begin{algorithmic}
    \caption{Pseudo-code for the entire algorithm}
    \label{alg: final}
    \State Given Some evolutionary operations $\mathcal{E}$ \Comment{i.e. CEM \cite{de_boer_tutorial_2005}}
    \State Given Some Initial buffer of randomly simulated designs $\mathcal{B}$
    \State Given reference index k \Comment{i.e. 10}
    \State Given DNN $\mathcal{M_{\theta}}$ parametrized by $\theta$
    \State update $\theta$ \Comment{i.e. 10 epochs}
    \While{num\_iter < max\_num\_iter}
        \State Get critical specification list $\mathcal{CS}$ according to the heuristic
        \State $\widetilde{\mathcal{B}} \leftarrow$ sort $\mathcal{B}$ by $cost(x)=\sum\limits_{i \in \mathcal{CS}} w_i*p_i(x) $
        \State $\mathcal{D}_{ref} = \widetilde{B}[k] $
        \State list of new children $\mathcal{L}$ = []
        \While{$\mathcal{L}$.length < 5} \Comment{i.e. until 5 children are approved}
            \State $\mathcal{D}_{new} \leftarrow\ \mathcal{E}.generate(\mathcal{B})$ \Comment{generate a new design}
            \State $\mathcal{P} \leftarrow M_{\theta}(\mathcal{D}_{new}, \mathcal{D}_{ref})$
            \If{$\mathcal{P}[i] = 1 ,\ \forall i \in \mathcal{CS}$}
                \State Run simulation on $\mathcal{D}_{new}$
                \State $\mathcal{L}$.append($\mathcal{D}_{new}$)
            \Else
                \State Continue
            \EndIf
        \EndWhile
        \State $\mathcal{B}\ \leftarrow\ \mathcal{E}$.select($\mathcal{B+L}$)
        \State update $\theta$ \Comment{i.e. 10 epochs}
    \EndWhile
    \end{algorithmic}
\end{algorithm}
\section{Experiments}
In this section we study a variety of experiments which clarifies some aspects of the algorithm and illustrates its capabilities on a variety of circuits. 
\subsection{Vanilla Two Stage OP-AMP}
\label{sec: ngspice}
First, to clarify the convergence behavior and benefits of the algorithm, we use a simple two stage op-amp evaluated only through schematic simulation using 45nm BSIM models on NGSPICE.
\par
\begin{figure}
\centering
\includegraphics[width=0.8\linewidth]{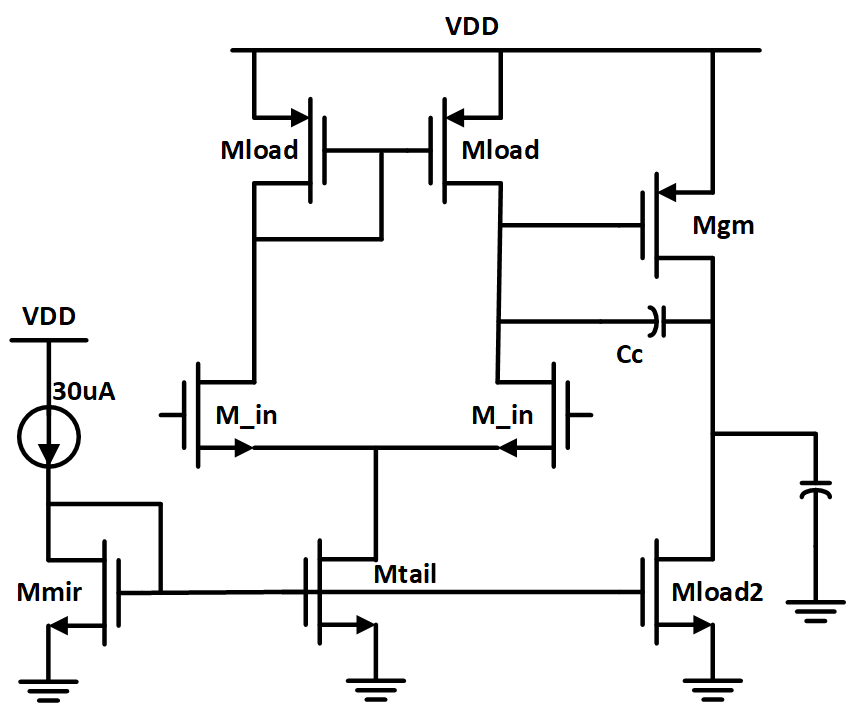}
\caption{Schematic of a vanilla two stage op-amp}
\label{fig:two_stage_ngspice}
\end{figure}

The circuit's schematic is shown in Figure \ref{fig:two_stage_ngspice}. The objective is to find the size of transistors and the value of the compensation capacitor such that the described circuit satisfies the requirements set in table \ref{tab: ngspice}. We fixed the length and width of the unit sized transistors to 45 $n$m and 0.5 $\mu$m, respectively, and for size of each transistor we limit the number of fingers to any integer number between 1 to 100. For compensation we also let the algorithm choose $C_c$ from any number between 0.1$p$F to 10$p$F with steps of 0.1$p$F. The grid size of the search space is $10^{14}$. A given instance is evaluated through DC, AC, CMRR, PSRR, and transient simulations, which in total takes one second for each design. Therefore, brute-force sweeping is not practical even in this simple example. However, This short simulation time allows us to do comparisons against the vanilla genetic algorithm and the oracle. Note that these methods are not feasible for layout-based simulations, since each RC extraction and simulation takes several minutes to run. 
\par

\begin{table}[]
\caption{Objectives of the design problem and performance of solutions found using different approaches for the two stage op-amp example}
\label{tab: ngspice}
\resizebox{\linewidth}{!}{%
\begin{tabular}{@{}lllll@{}}
\toprule
                            & Requirement      & Evolutionary & Oracle & \textbf{Ours} \\ \midrule
Gain                        & \textgreater 300 & 323          & 314    & \textbf{335}  \\
$f_{unity}$ {[}MHz{]}       & \textgreater 10  & 10.83        & 10.66  & \textbf{10.2} \\
Phase Margin {[}$^\circ${]} & \textgreater 60  & 60.7         & 60.83  & \textbf{62}   \\
$t_{settling}$ {[}ns{]}     & \textless 90     & 59.9         & 83.5   & \textbf{62}   \\
CMRR {[}dB{]}               & \textgreater 50  & 53           & 54     & \textbf{54}   \\
PSRR {[}dB{]}               & \textgreater 50  & 57           & 56     & \textbf{57}   \\
Systematic Offset {[}mV{]}  & \textless 1      & 0.823        & 0.94   & \textbf{0.32} \\
Ibias {[}$\mu A${]}         & \textless 200    & 188          & 158    & \textbf{148}  \\
\bottomrule
\end{tabular}%
}
\end{table}

Table \ref {tab: ngspice} shows the performance of the minimum cost solution found by different approaches. In this example, all approaches found a solution satisfying all specifications. It should be noted that the differences in the metrics are not due to the superiority of any of the algorithms, and is merely due to the stochastic nature of them.
\par

\begin{figure}
    \centering
    \begin{subfigure}{\linewidth}
        \centering
        \includegraphics[width=\textwidth]{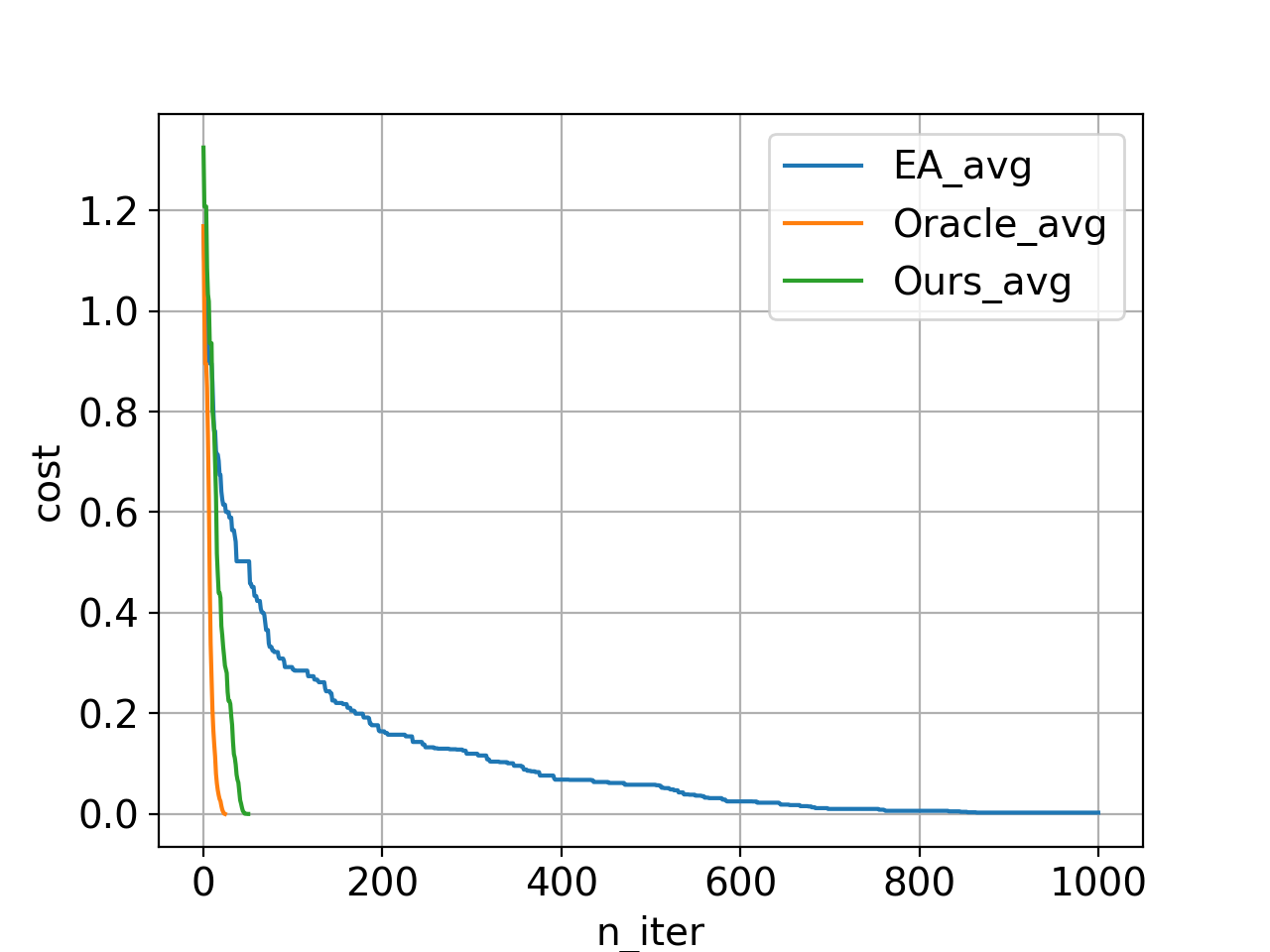}
        \caption{}
        \label{fig:iter}
    \end{subfigure}
    \vskip\baselineskip
    \vspace{-10pt}
    \begin{subfigure}{\linewidth}
        \centering
        \includegraphics[width=\linewidth]{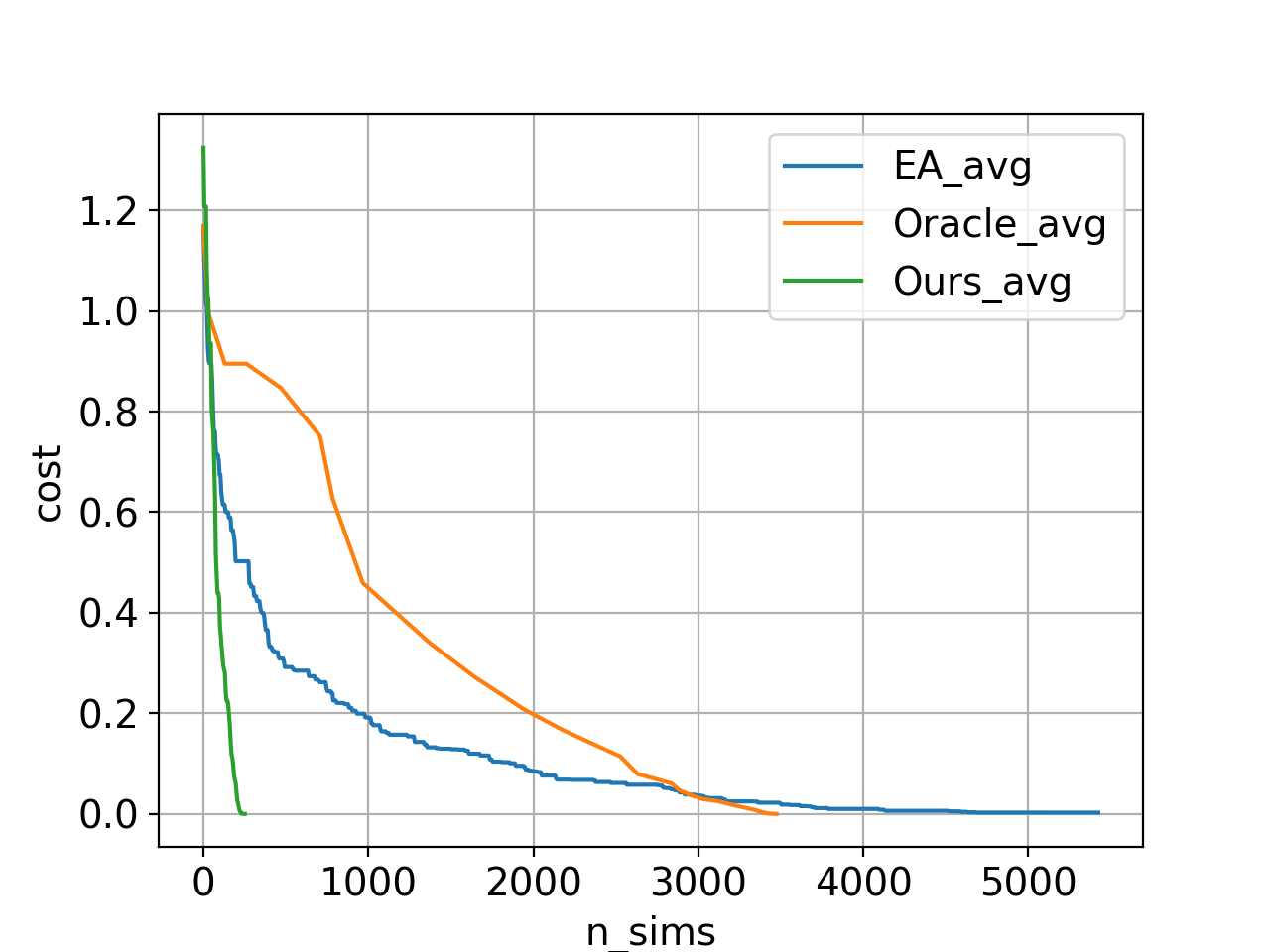}
        \caption{}
        \label{fig:sims}
    \end{subfigure}
    \caption{ a) Average cost of top 20 individuals across number of iterations. Each iteration corresponds to adding 5 designs to the population. b) Average cost of top 20 individuals across number of simulations.}
\end{figure}

Figure \ref{fig:iter} shows the average cost of the top 20 designs in the population for the oracle, evolutionary, and our algorithm. Reference design is always at rank 20 and each time only 5 designs are added. The evolutionary operations are also the same for all experiments. We ran our approach on multiple random initial seeds to ensure robustness in training and performance. We note that our approach significantly reduces the number of iterations (and also the number of simulations) compared to the same evolutionary algorithm. The gap shown between the oracle and our algorithm can be a measure of how much the prediction inaccuracy of the DNN is impacting the optimization process. The oracle runs simulation on all generated designs to make selection decisions. Therefore, it is impractical to run it on post-layout simulation on more complex circuits. In figure \ref{fig:sims} we show the same cost behavior for all three algorithms vs. number of simulations. The performance difference between the evolutionary and our approach is scaled with number of simulations per iteration. However, for the simulation-based oracle each iteration includes a variable number of simulation runs to find at least 5 better designs. Therefore, the gap between the oracle and our approach, in terms of number of simulations, is much more.
\par
Table \ref{tab: n_operation} shows a summary of number of operations in our simple example. We note that our approach can cut down a lot of simulations at the cost of more time spent on training and inference of a DNN. With recent advancements in hardware for machine learning and use of GPUs the time spent on training and inference can significantly be reduced. When we want to scale up to more intricate circuits there are two factors that makes our approach advantageous. First, when we do layout optimization, simulation drastically increases proportional to the circuit size. Moreover, complicated circuits have larger design space and it will become even more critical to prune out useless regions of design space as we scale up. 
\begin{table}[]
\caption{Summary of number of operations involved in the process of each approach}
\label{tab: n_operation}
\resizebox{\linewidth}{!}{%
\begin{tabular}{@{}llll@{}}
\toprule
                 & \# of NN Queries & \# of Re-training & \# of Simulations \\ \midrule
Simple Evolution & -                & -              & 5424              \\
Oracle           & -                & -              & 3474              \\
Ours             & 55102            & 50             & 241               \\ \bottomrule
\end{tabular}%
}
\end{table}


\subsection{Two Stage Op-Amp and Comparison with Expert Design}
\begin{figure}
    \centering
    \includegraphics[width=\linewidth]{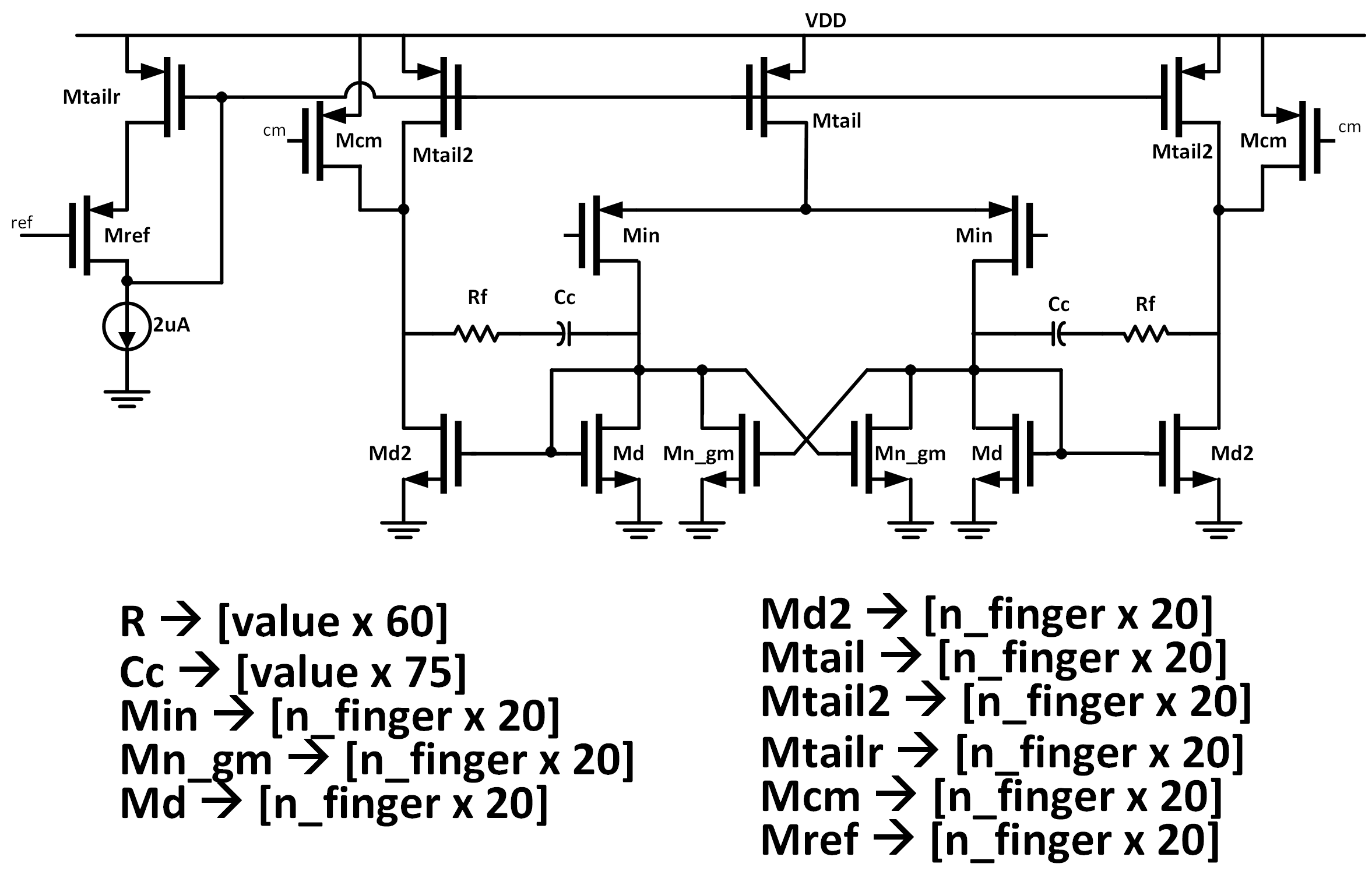}
    \caption{Two stage op-amp with negative $g_m$ load}
    \label{fig:bag_opamp}
\end{figure}

This example is presented to compare an expert-designed circuit with our algorithm's design. The op-amp's topology, and the cardinality of search space is shown in Figure \ref{fig:bag_opamp}, with each array denoting how many parameters were considered for design. For example for Mref, 20 values of n\_fingers were considered. In total, this design example has an 11 dimensional exploration space with size of $3\times10^{13}$. \par
The topology is a standard Miller compensated op-amp, in which the first stage contains diode-connected and negative-gm loads. The design procedure is more cumbersome than the previous two stage example, mainly because of the positive feedback. A scripted design procedure for this topology is included as part of BAG [\ref{fig:bag_opamp}] to exemplify codifying expert driven design methodologies. The design script is able to find the proper transistor sizing while considering layout parasitic effects using a closed loop design methodology. The inputs to the design script are specifications of phase margin and bandwidth, and the objective is to maximize gain. In this circuit the resistor and capacitor are schematic parameters, while all transistors and all connecting wires use the GF14 nm PDK extraction model. 
\par
Table \ref{tab: eric_opamp} shows a performance summary of our approach compared to that of the design generated by the design script. The script is unable to meet the gain requirement due to a designer-imposed constraint that the negative $g_m$ should not cancel more than 70\% of the total positive resistance at first stage's output. This constraint arises from a practical assumption that there will be mismatch between the negative $g_m$'s resistance and the overall positive resistance, due to PVT variations. Thus, the circuit can become unintentionally unstable, and therefore during design we leave some margin to accommodate these prospective random variations. We have the option of imposing a similar constraint to equate the design spaces, or we can run simulations over process and temperature variations to ensure that our practical constraints are not too pessimistic.
\par 
For our approach, the initial random population size is 100 with a best cost of 0.3. We ran every simulation on different PVT variations, and recorded the worst metric as the overall performance value. 
Reaching a solution with our approach took 3 hours including initial population characterization, whereas developing the design script takes 4-7 days according to the expert. The DNN was queried 3117 times in total (equivalent to 6 minutes of run time on our compute servers) and we only ran 120 new simulations in addition to the initial population of size 100 (each of which takes on average 48 seconds to run). Moreover, the complexity of developing a design script forces the designer to limit the search space in an effort to make the process feasible and a generic design algorithm that properly imposes high-level specifications onto a large system is immensely difficult to generate. We will see an example of such systems and our approach's solution in the next section.
\par

\begin{table}[]
\caption{Performance of expert design methodology and our approach}
\label{tab: eric_opamp}
\resizebox{0.8\linewidth}{!}{%
\begin{tabular}{@{}llll@{}}
\toprule
             & Requirement                 & Expert       & Ours         \\ \midrule
$f_{unity}$ & \textgreater 100 MHz        & 382 MHz      & 159 MHz      \\
pm           & \textgreater{} $60^{\circ}$ & $64^{\circ}$ & $75^{\circ}$ \\
gain         & \textgreater 100 (for ours) & 42           & 105          \\ \bottomrule
\end{tabular}%
}
\end{table}

\subsection{Full optical link receiver}
\begin{figure}
    \centering
    \includegraphics[width=\linewidth]{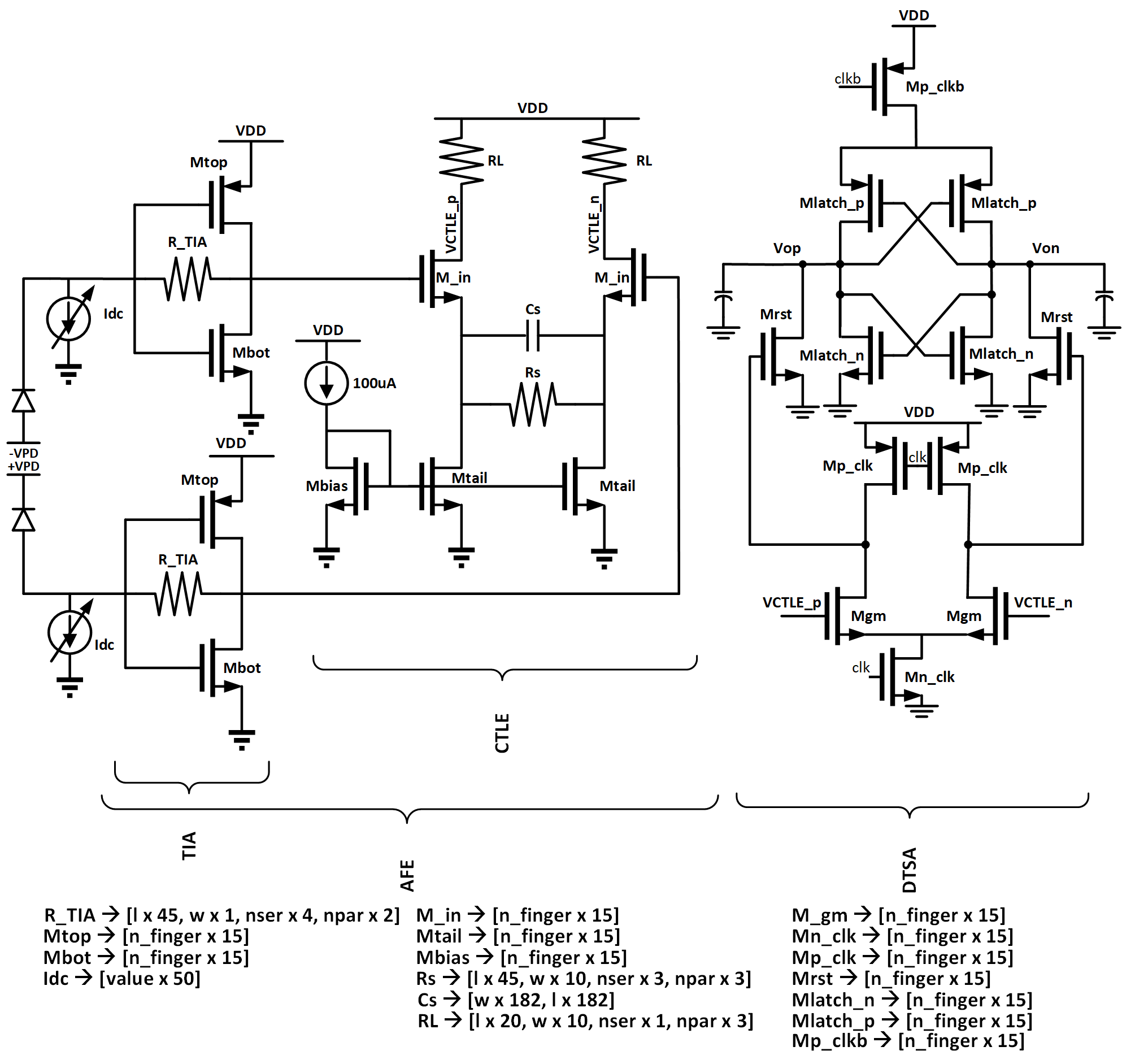}
    \caption{Optical receiver schematic}
    \label{fig:sch_rx}
\end{figure}

The following experiment highlights the capabilities of our approach in handling complex analog/mixed signal design problems using post-layout simulations. We demonstrate a differential optical link receiver front-end with a one tap double tail sense amplifier (DTSA) in the end. The circuit is shown in Figure \ref{fig:sch_rx} with design space parameters at the bottom. The goal is to design this circuit from very high level specifications, namely, data rate, power consumption, and minimum sensitivity for a given bit error rate (BER). 
\par

\begin{figure}
    \centering
    \begin{subfigure}{\linewidth}
        \centering
        \includegraphics[width=\textwidth]{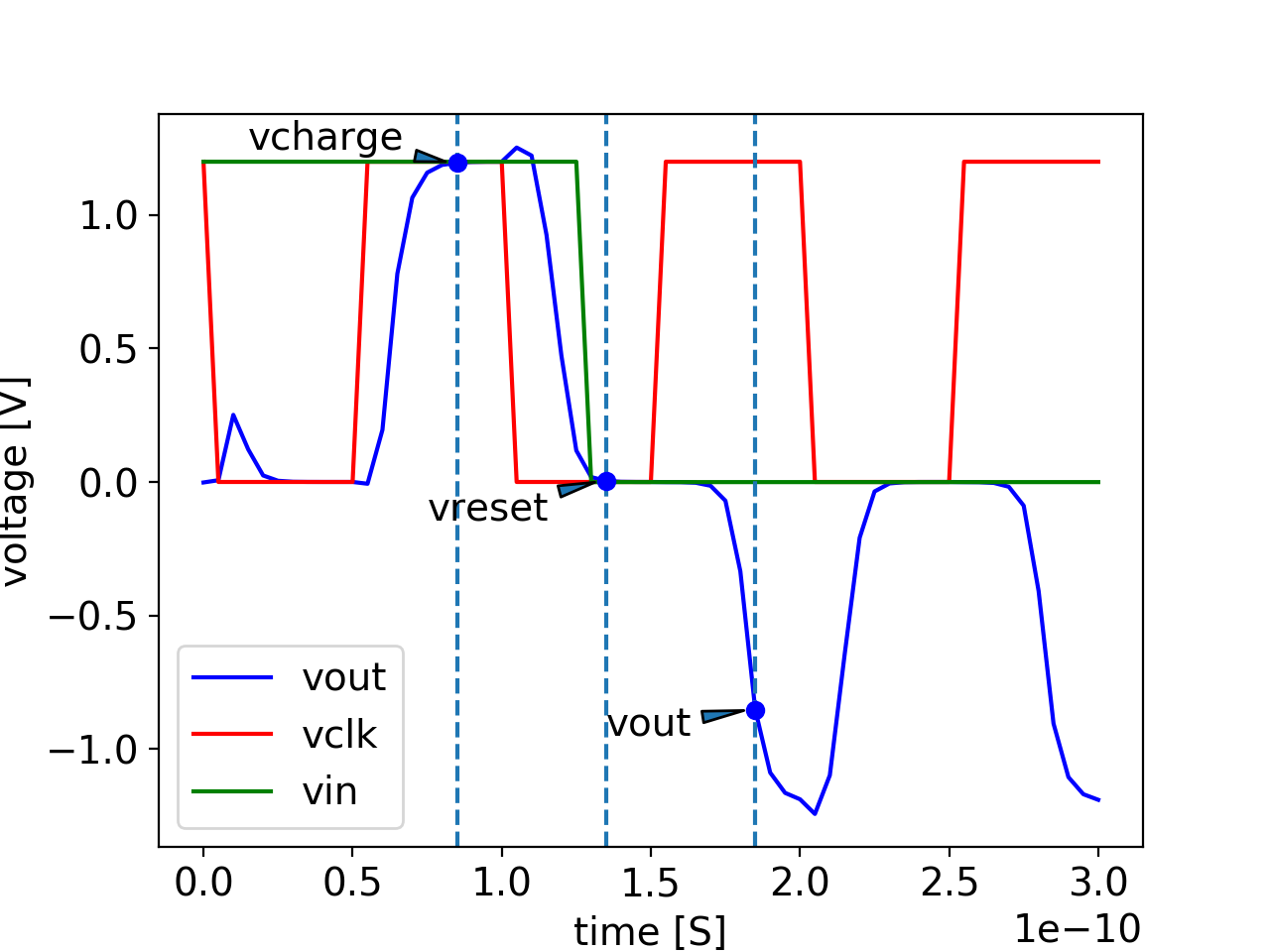}
        \caption{}
        \label{fig:sims_od}
    \end{subfigure}
    \vskip\baselineskip
    \begin{subfigure}{\linewidth}
        \centering
        \includegraphics[width=\linewidth]{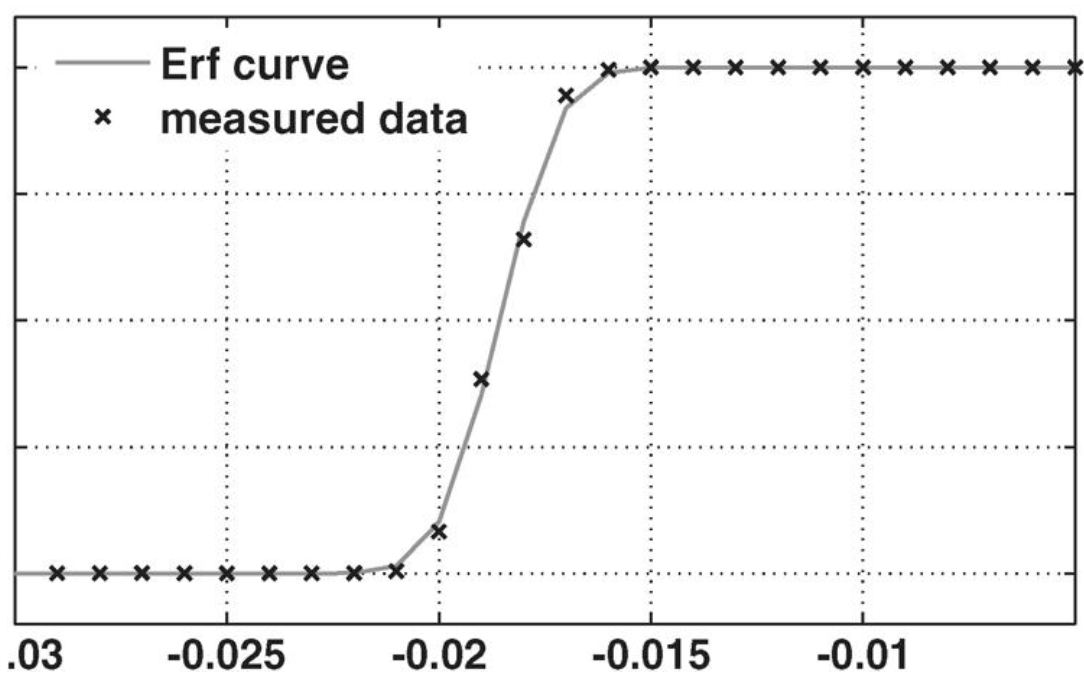}
        \caption{}
        \label{fig:sims_erf}
    \end{subfigure}
    \caption{a) Overdrive test recovery simulation curves for DTSA b) Probability of outputting a one vs. $V_{in}$. We can use the cumulative density function of a Gaussian to estimate the standard deviation of the noise}
\end{figure}

\begin{figure}
    \centering
    \begin{subfigure}{\linewidth}
        \centering
        \includegraphics[width=\linewidth]{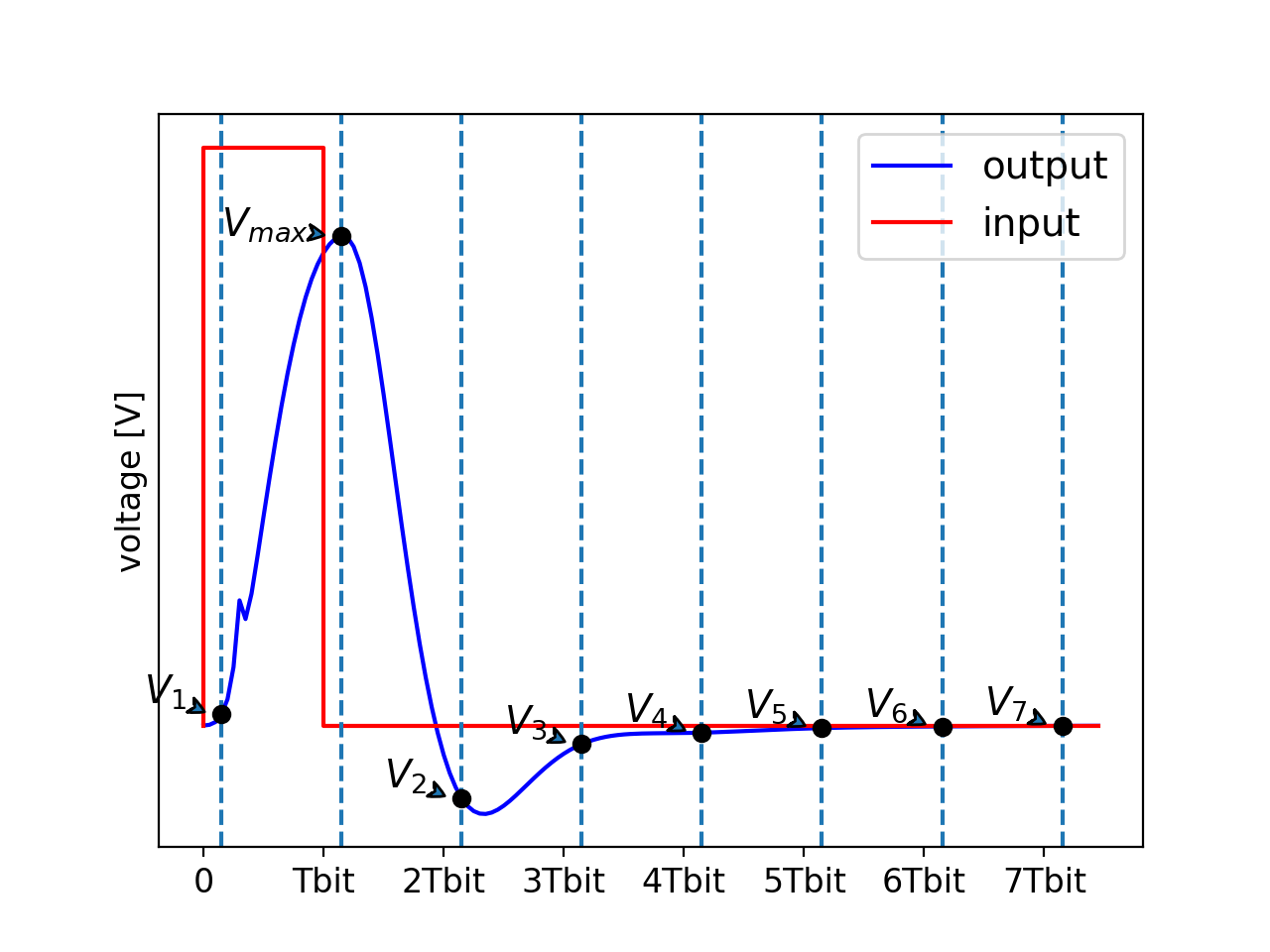}
        \caption{}
        \label{fig:sims_eye_est}
    \end{subfigure}
    \vskip\baselineskip
    \vspace{-13pt}
    \begin{subfigure}{\linewidth}
        \centering
        \includegraphics[width=0.8\linewidth]{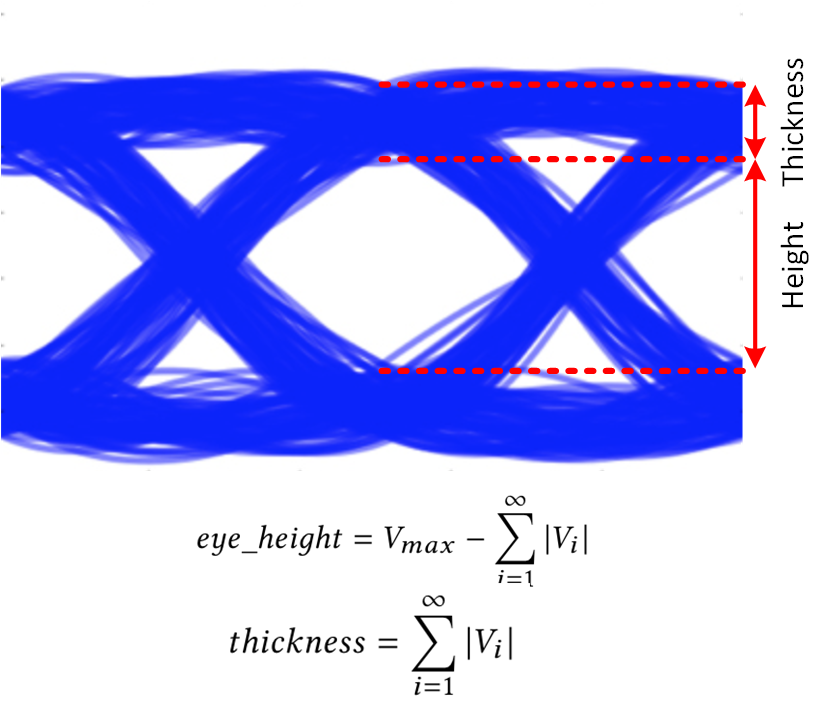}
        \caption{}
        \label{fig:sims_eye_calc}
    \end{subfigure}
    \caption{a) Input and output signals for measuring the eye height and thickness, the input is a small signal pulse with an amplitude of target input sensitivity, and with a width of $T_{bit}$ for the target data rate. The output curve is sampled at time instances shown relative to the main cursor (maximum of output) b) Equations used for estimating eye height and thickness to express the fidelity of eye diagram}
\end{figure}

Automating characterization of instances of this circuit is the key in setting up the environment prior to running the algorithm. The following steps are cruicial to get performance metrics on each design. First we instantiate the DTSA's layout, schematic, and extracted netlist. We then run overdrive test recovery to characterize the transient behaviour of DTSA. Figure \ref{fig:sims_od} shows a typical overdrive test recovery curve for a given comparator. We specifically measure \textit{vcharge}, \textit{vreset}, and \textit{vout} in the time instances relative to the edge of the clock as shown in the figure. By specifying these three numbers as well as a minimum \textit{vin} (i.e 1 mV), we can describe the performance of the comparator at a given data rate. To get the noise behaviour of the comparator, we run numerous transient noise simulations for several cycles while sweeping input voltage from a small negative number to a small positive number. We can then fit a normal Gaussian distribution to the estimated probability of ones in each transient run and get an estimation of the input referred voltage noise of the DTSA. Figure \ref{fig:sims_erf} illustrates this simulation procedure. We then take the entire system's extracted netlist and characterize the behaviour of the analog front end (AFE) while the DTSA is acting as a load for the continuous time linear equalizer (CTLE). 
\par
Once we used noise simulations to get the input referred voltage noise of the comparator, we can then aggregate the comparator's noise from previous simulations with the AFE's noise to compute the total rms noise comparator's input. We also use the transient response of the circuit to assure high fidelity for the eye diagram at the input of comparator. The input/output curves, and formulas used to measure eye's fidelity are shown in Figures \ref{fig:sims_eye_est} and \ref{fig:sims_eye_calc}. 
\par
We estimate eye height and eye thickness ratio, and specify a constraint on them to describe the quality of eye diagram for a given input sensitivity.
Using BER of $10^{-12}$ we can compute the required eye height at the input of the comparator using equation \ref{eq: eye_h} and compare it against the actual eye height. For the optimization objective we can put a constraint on the relative difference of the actual eye height and the required eye height (i.e. actual eye height should be 10\% larger than the minimum required eye height. We call this percentage eye margin).\par
\begin{equation}
    BER = 0.5*\mathnormal{erfc}(\frac{eye\_h_{min} - \text{Residual Offset} - DTSA_{sens.}}{\sqrt{2}\sigma_{noise}})
    \label{eq: eye_h}
\end{equation}


For sensitivity we use the $V_{in}$ from the overdrive test recovery in previous tests. There will be also some mismatch offset which can significantly be reduced with a systematic offset cancellation scheme. The offset cannot, however, be fully eliminated, so the residual offset will also be considered (i.e. 1mV).
We also run a common mode ac simulation to ensure that the tail transistors providing bias currents are operating in saturation by specifying a reasonable minimum common mode rejection ratio requirement. For one instance, this whole process takes about 200 seconds on our compute servers. We can then compute the cost of each design as specified by equation \ref{eq: cost}. 
\par
In terms of layout generator search space, each resistor drawn in Figure \ref{fig:sch_rx} has unit length, unit width, number of series units, and number of parallel units. The CTLE's capacitor has width and length, and each transistor has number of fins and number of fingers that need to be determined. We fixed the number of fins to simplify the search space. The cardinality of each design parameter is written in Figure \ref{fig:sch_rx}. In total, the design example has a 26 dimensional exploration space with size of $2.8\times10^{30}$.
\par

\begin{table}
\caption{Design Performance for optical receiver design for $C_{PD} = 20 fF$, $I_{min}=3\mu A, \text{Data Rate } = 10 Gbit/s$}
\label{tab: afe_results}
\resizebox{\linewidth}{!}{%
\begin{tabular}{lllll}
\hline
                                                               & Requirement          & \multicolumn{1}{c}{\begin{tabular}[c]{@{}c@{}}A \\ (cost = 0)\end{tabular}} & \multicolumn{1}{c}{\begin{tabular}[c]{@{}c@{}}B \\ (cost = 0.1)\end{tabular}} & \multicolumn{1}{c}{\begin{tabular}[c]{@{}c@{}}C \\ (cost = 0.2)\end{tabular}} \\ \hline
Thickness ratio                                                & \textless 10\%       & 6\%                                                                       & 3.3\%                                                                        & 8.5\%                                                                         \\
\begin{tabular}[c]{@{}l@{}}Eye margin\\ @ $I_{min}$\end{tabular} & \textgreater 10\%    & 10.2\%                                                                        & 15.1\%                                                                         & 10.4\%                                                                         \\
CMRR                                                           & \textgreater 3       & 4.77                                                                         & 5.51                                                                          & 4.5                                                                             \\
vcharge                                                        & \textgreater 0.95VDD & VDD                                                                         & VDD                                                                       & VDD                                                                           \\
vreset                                                         & \textless 1 mV       & 20 $\mu V$                                                                  & 52 $\mu V$                                                                    & 21 $\mu V$                                                                    \\
vout                                                           & \textless -0.9VDD    & -0.91VDD                                                                    & -0.88VDD                                                                       & -0.87VDD                                                                          \\
Total Noise                                                    & \textless 5 mV       & 2.9 mV                                                                      & 3.2 mV                                                                        & 3.05 mV                                                                        \\
Total Ibias                                                    & \textless 10 mA      & 6.2 mA                                                                      & 4.04 mA                                                                       & 6.3 mA                                                                        \\ \hline
\end{tabular}%
}
\end{table}

\begin{figure}
    \centering
    \includegraphics[width=\linewidth]{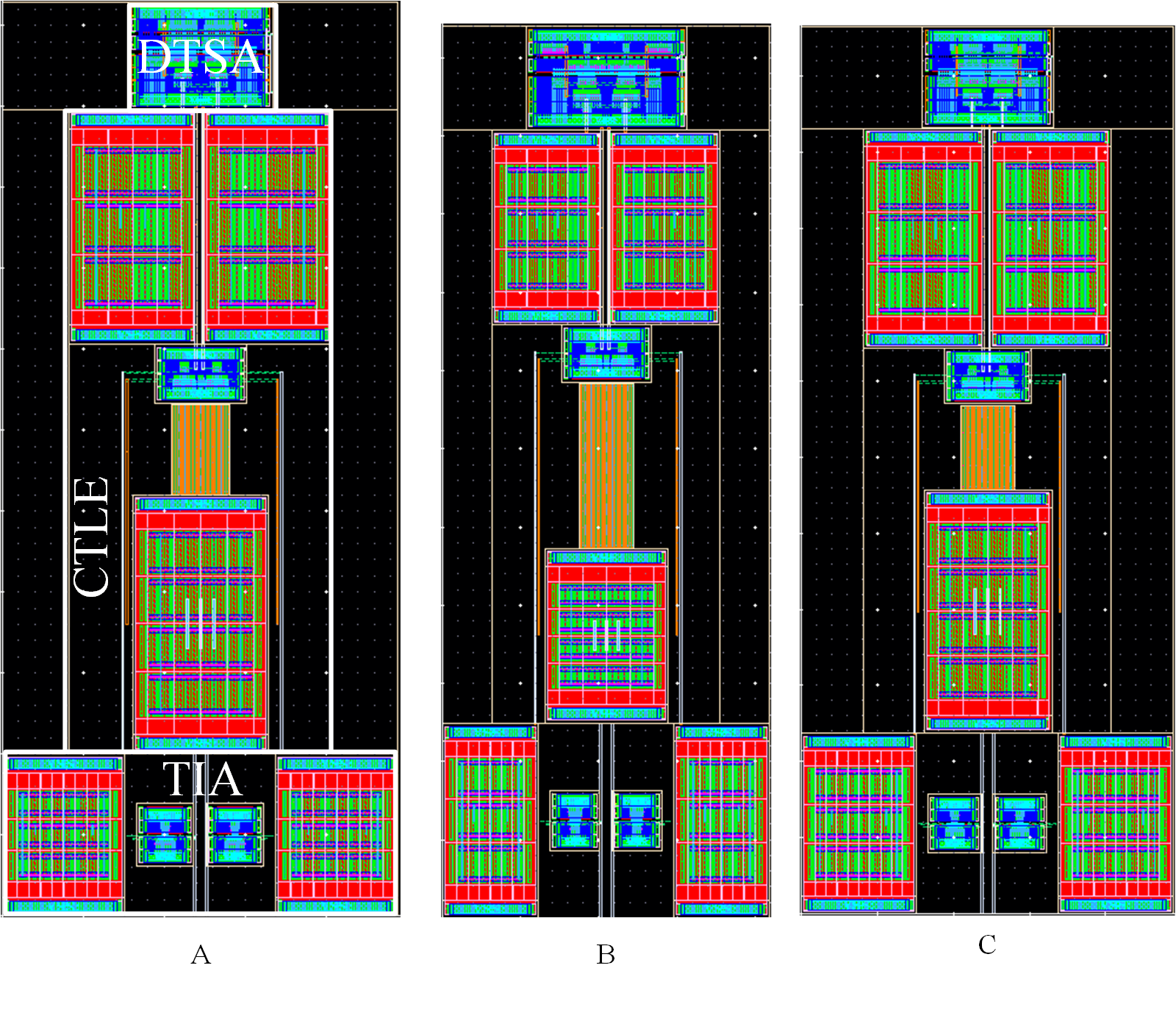}
    \caption{Sample layouts of the optical link receiver circuit with different cost values of 0, 0.1, and 0.2 for A, B, and C, respectively.}
    \label{fig:layout_costs}
\end{figure}

Figure \ref{fig:layout_costs} and Table \ref{tab: afe_results} show the layout and performance of the solution for designs found with cost of 0 (satisfying all specs), 0.1, and 0.2, respectively. The first design solution was found using 435 simulations equivalent to 27 hours of run time. This number includes generating the initial population which consisted of 150 designs with the best cost function of 2.5. During the process the DNN was queried 77487 times in total and only 285 of those were simulated, representing around 300x sample compression efficiency. From the total run time, 1.6 hours were spent on training and almost 2.1 hours were spent on querying the DNN.

\subsection{Conclusion}
In this paper we have proposed a new sample-efficient evolutionary-based optimization algorithm for designing analog circuits using analog layout generators. In this approach we used DNNs to prune-out the useless part of design space, so that we can save time on long simulations. We showed that the algorithm can be used in designing a variety of real, practical analog/mixed signal circuits with different applications regardless of size and complexity, as long as the verification procedure is properly defined. 
\vspace{50pt}

\bibliographystyle{acm}

\bibliography{bib}

\end{document}